\documentclass[preprintnumbers,superscriptaddress,showkeys,showpacs,byrevtex]{revtex4}   
\usepackage{amsmath,amsfonts,amssymb,amscd,amsxtra,amsthm}
\usepackage{graphicx}
\usepackage{bm}
\usepackage{epstopdf}
\usepackage{multirow}
\begin{document}  
\preprint{INHA-NTG-11/2011}
\preprint{KIAS-P11049}
\title{$K^*\Lambda(1116)$ photoproduction and nucleon resonances}      
\author{Sang-Ho Kim} 
\email[E-mail: ]{sanghokim@inha.edu}
\affiliation{Department of Physics, Inha University, Incheon 402-751, Republic of  Korea}
\author{Seung-il Nam}
\email[E-mail: ]{sinam@kias.re.kr}
\affiliation{School of Physics, Korea Institute for Advanced Study,
  Seoul 130-722, Republic of Korea} 
\author{ Yongseok Oh }
\email[E-mail: ]{yohphy@knu.ac.kr}
\affiliation{School of Physics and Energy Sciences, Kyungpook National
  University,  Daegu 702-701,  Republic of Korea } 
\author{Hyun-Chul Kim} 
\email[E-mail: ]{hchkim@inha.ac.kr}
\affiliation{Department of Physics, Inha University, Incheon 402-751, Republic of  Korea} 
\date{\today}
\begin{abstract}
We study the reaction mechanism of $K^*\Lambda(1116)$ photoproduction
off the proton target near threshold considering the contributions
from nucleon-resonances.  
Employing the effective Lagrangian method at the tree-level Born
approximation, we investigate the role of the $D_{13}$(2080) and the
$D_{15}$(2200). We found that the $D_{13}$ plays a crucial role to
reproduce the experimental data well in the forward scattering region
at low energy. We also present theoretical predictions on the single
photon-beam asymmetry ($\Sigma$) as well as the energy and angular
dependence for the cross sections of this reaction. 
\end{abstract}  
\pacs{13.75.Cs, 14.20.-c}
\keywords{$K^*$ photoproduction, effective Lagrangian method, nucleon
  resonances}    
\maketitle
\section{Introduction}
Strange-meson photoproduction is one of the most practical and useful
experimental and theoretical methods to investigate the strangeness
production processes in hadron physics.  
In this article, we report our recent study for $K^*\Lambda(1116)$
photoproduction, which employs the effective Lagrangian method in the
tree-level Born approximation. In previous studies, it was noted that
the production rate of theoretical calculations is insufficient to
reproduce the experimental data for this process in the threshold
energy
region~\cite{Guo:2006kt,Hicks:2010pg,CLASKSTAR,Oh:2006hm,Oh:2006in}.  
To explain the discrepancies between theory and experiments, we
investigate the role of nucleon resonances whose masses are in this
energy region. For this purpose, we consider the $D_{13}(2080)$ and
the $D_{15}(2200)$ nucleon resonances on top of the relevant nucleon
Born terms. We use the resonance parameters extracted from 
experimental information of the PDG~\cite{Nakamura:2010zzi} or the
theoretical predictions of the relativistic quark-model of
Refs.~\cite{Capstick:1992uc,Capstick:1998uh}. Our numerical results
show that the role of the $D_{13}(2080)$ is essential to reproduce the
experimental data for the cross sections of $K^*\Lambda(1116)$
photoproduction, whereas the $D_{15}(2200)$ contribution is rather
small. This implies that the nucleon resonance contribution is crucial
in the production mechanisms of $K^*\Lambda$ photoproduction near the
threshold. The present article is organized as follow. In Section 2,
we briefly introduce the theoretical framework for resonance
contributions. The numerical results are presented in Section 3 with
discussions. Section 4 contains a summary and conclusion. 
\section{Theoretical Formalisms}
\begin{figure}[t]
\centering
\includegraphics[width=12cm]{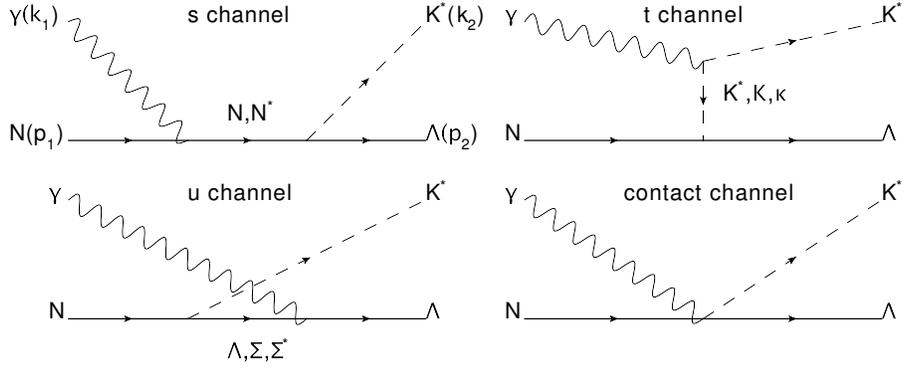}
\caption{Relevant Feynman diagrams for $K^*\Lambda(1116)$
  photoproduction at tree level.  
All the momenta for the involved particles are also defined.}       
\label{FIG1}
\end{figure}

For studying the production mechanisms of $K^*\Lambda(1116)$
photoproduction, we consider the Feynman diagrams as depicted in
Fig.~\ref{FIG1}. Referring the details on the theoretical formalism
for the nucleon Born terms and the form factor prescription to
Refs.~\cite{Oh:2006hm,Oh:2006in}, we focus on the role of nucleon
resonances. Here, we consider the $D_{13}(2080,3/2^-)$ and the
$D_{15}(2200,5/2^-)$ whose masses are close to the threshold energy of
$\gamma N \to K^* \Lambda(1116)$. The effective Lagrangians for the
electromagnetic (EM) interactions are 
\begin{eqnarray}
\label{eq:LAGEM}
\mathcal{L}_{\gamma  ND_{13}}
&=& -\frac{ieh_{1D_{13}}}{2M_N} \bar N  \gamma_\nu F^{\mu\nu} R_\mu
-\frac{eh_{2D_{13}}}{(2M_N)^2} \partial_\nu \bar N  F^{\mu\nu} R_\mu 
+ \mathrm{h.c.}                       
\cr
\mathcal{L}_{\gamma ND_{15}} 
&=& \frac{eh_{1D_{15}}}{(2M_N)^2} \bar N  \gamma_\nu \gamma_5 \partial^\alpha 
F^{\mu\nu} R_{\mu\alpha}-\frac{ieh_{2D_{15}}}{(2M_N)^3} \partial_\nu
\bar N \gamma_5 \partial^\alpha F^{\mu\nu} R_{\mu\alpha} +
\mathrm{h.c.},  
\end{eqnarray}
where $R$ stands for the resonance field with given spin and parity. 
The strengths of the couplings in Eq.~(\ref{eq:LAGEM}) can be
determined from the experimental values for the corresponding helicity
amplitudes~\cite{Nakamura:2010zzi} or quark model \cite {Capstick:1992uc},
which gives $h_{1D_{13}}=0.829$,$h_{2D_{13}}=-0.845$,  $h_{1D_{15}}=0.346$, and $h_{2D_{15}}=0.031$. 

The effective Lagrangians for the relevant interactions of the
resonances with the $K^*\Lambda$ channel are 
\begin{eqnarray}
\label{eq:STRONG}
\mathcal{L}_{K^* D_{13} \Lambda }
&=& -\frac{i}{2M_N} \left[ g_{1D_{13}} \bar \Lambda \gamma_\nu
-\frac{ig_{2D_{13}}}{2M_N} \partial_\nu \bar \Lambda                                          
+\frac{ig_{3D_{13}}}{2M_N} \bar \Lambda
\partial_\nu  \right] K^{*\mu\nu} R_\mu + \mathrm{h.c.}      
\cr
\mathcal{L}_{K^* D_{15}\Lambda }
&=& \frac{1}{(2M_N)^2}
\left[g_{1D_{15}} \bar \Lambda \gamma_\nu \gamma_5
 \partial^\alpha  -\frac{ig_{2D_{15}}}{2M_N} \partial_\nu \bar \Lambda \gamma_5
\partial^\alpha +\frac{ig_{3D_{15}}}{2M_N} 
\bar \Lambda \gamma_5 \partial^\alpha \partial_\nu  \right]K^{*\mu\nu} R_{\mu\alpha}
 + \mathrm{h.c.}.
\label{eq:strong}
\end{eqnarray}
The strong coupling constants in Eq.~(\ref{eq:STRONG}) are unknown
and, to fix these parameters, we use the theoretical estimations in
the relativistic quark model~\cite{Capstick:1998uh} on
the partial decay widths of the resonances, 
\begin{equation}
\label{eq:GGG}
\Gamma_{R\to K^*\Lambda}=\sum_\ell|G(\ell)|^2\,,
\end{equation}
where the values for $G(\ell)$ are given in
Refs.~\cite{Capstick:1998uh}. The partial decay width
$\Gamma_{R\to K^*\Lambda}$ can be calculated from the interaction
Lagrangians in Eq.~(\ref{eq:STRONG}) and we can estimate the strong
coupling constants of Eq.~(\ref{eq:strong}). 
Since we focus on the role of nucleon resonances near the threshold, we
take into account only the lowest partial-wave contributions.  
Hence, we neglect the $g_{(2,3)R}$ terms in Eq.~(\ref{eq:STRONG}) and
employ only the contributions of the lowest orbital angular momentum
in $G(\ell)$. With this simplified model, we present the numerical
results using $|g_{1D_{13}}^{}|=1.59$ and $|g_{1D_{15}}^{}|=1.03$.  
The relative signs of these coupling constants to the other
contributions will be examined by comparing with the experimental data
on the cross sections. All the other couplings and parameters are the
same as in Refs.~\cite{Oh:2006hm,Oh:2006in}.  
\section{Numerical results}
\begin{figure}[t]
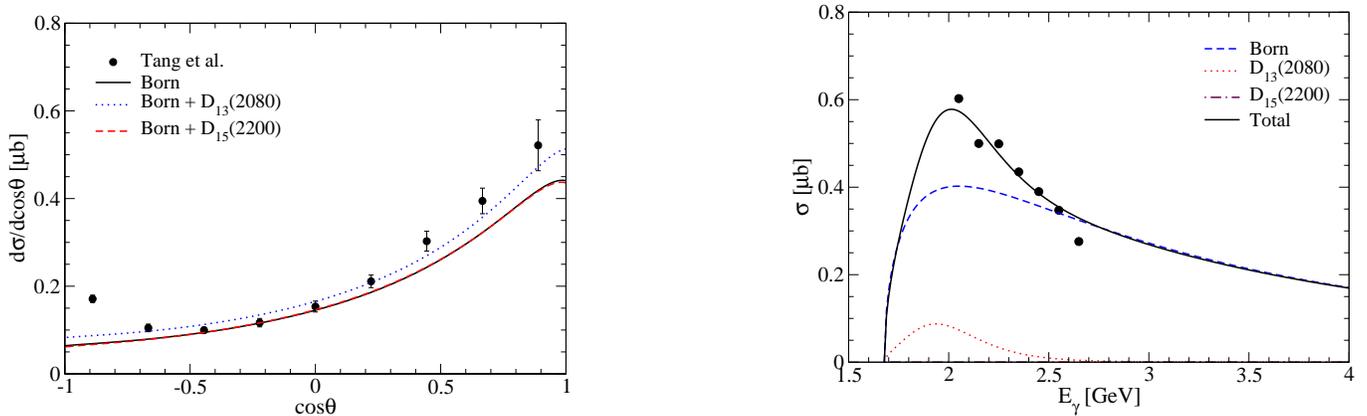

\bigskip\smallskip
\includegraphics[width=7.5cm]{FIG3.eps} \hfill
\includegraphics[width=7.5cm]{FIG2.eps}
\caption{(Color online) Differential cross sections at $E_\gamma=2.35$
  GeV as a function of $\cos\theta$ (left) and  
the total cross sections as a function of $E_\gamma$ (right) with and
without the nucleon resonances.  
The experimental data are taken from Ref.~\cite{Hicks:2010pg}.}       
\label{FIG23}
\end{figure}

In this Section, we present the numerical results for
$K^*\Lambda(1116)$ photoproduction off the proton target.  
In the left panel of Fig.~\ref{FIG23}, our results for the
differential cross sections $d\sigma/d\cos\theta$ for this reaction 
are shown with respect to $\cos\theta$, where $\theta$ denotes the
scattering angle between the incident photon beam and outgoing $K^*$
in the center-of-mass frame. 
Here, we only present the results for $E_\gamma=2.35$ GeV.  
This shows that the role of the resonances is crucial to reproduce the
recent experimental data from the CLAS Collaboration at
TJNAF~\cite{Hicks:2010pg}. 
Furthermore, we found that the $D_{15}$ contribution is minor in
comparison to that of the $D_{13}$. Note that our results
underestimate the data in the backward scattering region, i.e. 
$\cos\theta\approx-1$.  
We ascribe this discrepancy to underestimating the role of the
$u$-channel hyperon-pole contributions. 
In the present study, we do not consider $u$-channel hyperon
resonances apart from the $\Lambda$, the $\Sigma$, and the $\Sigma^*$.  
Varying the photon beam energy, we verified that the experimental
data for the differential cross sections 
in the range of $E_\gamma=(2.15\sim2.65)$ GeV are qualitatively well
reproduced in this model. The details will be reported elsewhere.

Shown in the right panel of Fig.~\ref{FIG23} are the results for the
total cross sections with and without the resonance contributions.
The experimental data are extracted from those of the differential
cross sections for $E_\gamma=(2.05\sim2.65)$ GeV in
Ref.~\cite{Hicks:2010pg}, and we only quote the central values.  
Here, the nucleon Born term contribution is given by the dashed curve,
while the $D_{13}$ and $D_{15}$ contributions are drawn in the dotted
and the dot-dashed ones, respectively. This again shows that the
$D_{15}$ contribution is negligible, and the production rate in the
energy of $E_\gamma \le 2.5$~GeV is largely dominated by the $D_{13}$
contribution.   
The enhancement of cross sections in this energy region can not be
explained in the simple Born-term calculations in
Refs.~\cite{Oh:2006hm,Oh:2006in,Kim:2011ph}.  
We thus conclude that the resonance contribution, in particular the
$D_{13}$ contribution, is crucial in understanding the mechanism of
the $K^*\Lambda(1116)$ photoproduction.  

\begin{figure}[t]
\centering \bigskip\smallskip
\includegraphics[width=7.5cm]{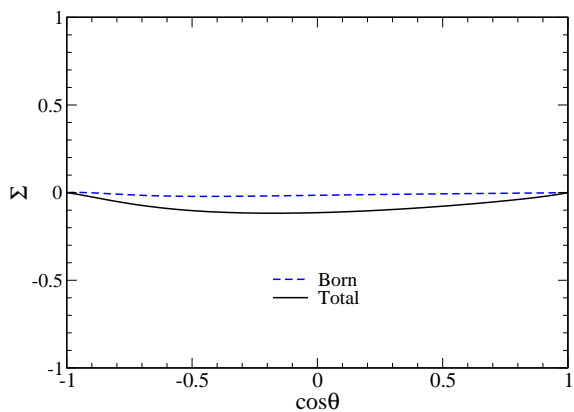}
\caption{(Color online) Photon-beam asymmetry $\Sigma$ with and
  without the nucleon-resonance contributions as functions of
  $\cos\theta$ at $E_\gamma=2.35$ GeV.}        
\label{FIG4}
\end{figure}

Finally, we compute a single-polarization quantity, i.e. the
photon-beam asymmetry, defined as  
\begin{equation}
\label{eq:BA}
\Sigma=\frac{d\sigma_\parallel-d\sigma_\perp}{d\sigma_\parallel+d\sigma_\perp},
\end{equation}
where the subscripts $\perp$ and $\parallel$ stand for the photon-beam
polarization being perpendicular and parallel, respectively, to the
reaction plane. In Fig.~\ref{FIG4}, we show the result for $\Sigma$ at
$E_\gamma=2.35$ GeV. This shows that $\Sigma$ is almost zero without
nucleon resonances, and the inclusion of the resonance contributions,
which are dominated by the $D_{13}$, can give non-vanishing asymmetry
in the region of $\cos\theta\approx(-0.15\sim0.5)$, which indicates
that the magnetic-transition contribution becomes larger.  

\section{Summary and Conclusion}
In this work, we investigated the mechanism of the $K^*\Lambda$
photoproduction based on the effective Lagrangian method at the
tree-level Born approximation with nucleon resonances. 
Our results are obtained with the parameters extracted from 
presently available experimental and theoretical information  
for the $D_{13}$ and $D_{15}$ resonances. We found that the
$D_{13}(2080)$ resonance contribution is important and has a crucial
role to bring the theoretical results close to the experimental data, 
in particular, near the threshold region, while the $D_{15}$
contribution is negligible. More detailed analyses with the results on
various physical observables are under progress and will be reported
elsewhere. 
\begin{acknowledgements}
We are grateful to K.~Hicks and A.~Hosaka for constructive comments 
and fruitful discussions. Two of us (S.H.K. and H.Ch.K.) were
supported by Basic Science Research Program through the National
Research Foundation of Korea (NRF) funded by the Ministry of
Education, Science and Technology (Grant \mbox{No.} 2009-0089525). The
work of S.i.N. was partially supported by the Grant NRF-2010-0013279
from the NRF of Korea, and Y.O. was supported by Basic Science
Research Program through the National Research Foundation of Korea
(NRF) funded by the Ministry of Education, Science and Technology
(Grant \mbox{No.} 2010-0009381).
\end{acknowledgements}

\end{document}